\documentclass[superscriptaddress,nofootinbib,10pt,preprintnumbers]{revtex4}
\usepackage{graphicx,amssymb,amsmath,slashed}
\usepackage{hyperref}

\newcommand{\beq}{\begin{equation}}
\newcommand{\eeq}{\end{equation}}
\newcommand{\co}{\mathcal{O}}

\newcommand{\tr}{\mathrm{Tr}}
\newcommand{\yut}{(Y_u Y_u^\dagger)_{\slashed{\mathrm{tr}}}}
\newcommand{\ydt}{(Y_d Y_d^\dagger)_{\slashed{\mathrm{tr}}}}
\newcommand{\au}{\mathcal{A}_u}
\newcommand{\ad}{\mathcal{A}_d}
\newcommand{\im}{\mathrm{Im}}
\newcommand{\re}{\mathrm{Re}}
\newcommand{\br}{\mathrm{Br}}

\begin{document}

\preprint{\tiny CERN-PH-TH/2012-030}

\title{\boldmath On the Universality of CP Violation in $\Delta F=1$ Processes}

\author{Oram Gedalia}
\affiliation{\mbox{Department of Particle Physics and
Astrophysics, Weizmann Institute of Science, Rehovot 76100,
Israel}}

\author{Jernej F.\ Kamenik}
\affiliation{J. Stefan Institute, Jamova 39, P.O. Box 3000,
1001 Ljubljana, Slovenia} \affiliation{Department of Physics,
University of Ljubljana, Jadranska 19, 1000 Ljubljana,
Slovenia}

\author{Zoltan Ligeti}
\affiliation{Ernest Orlando Lawrence Berkeley National
Laboratory, University of California, Berkeley, CA 94720,
United States}

\author{Gilad Perez}
\affiliation{\mbox{Department of Particle Physics and
Astrophysics, Weizmann Institute of Science, Rehovot 76100,
Israel}} \affiliation{CERN, Theory Division, CH1211 Geneva 23,
Switzerland}

\begin{abstract}

We show that new physics which breaks the left-handed $SU(3)_Q$
quark flavor symmetry induces contributions to CP violation in
$\Delta F=1$ couplings which are approximately {\it universal},
in that they are not affected by flavor rotations between the
up and the down mass bases. (Only the~short distance
contributions are universal, while observables are also
affected by hadronic matrix elements.) Therefore, such flavor
violation cannot be aligned, and is constrained by the
strongest bound from either the up or the down sectors. We use
this result to show that the bound from $\epsilon'/\epsilon$
prohibits an $SU(3)_Q$ breaking explanation of the recent LHCb
evidence for CP violation in $D$ meson decays. Another
consequence of this universality is that supersymmetric
alignment models with a moderate mediation scale are consistent
with the data, and are harder to probe via CP violating
observables. With current constraints, therefore, squarks need
not be degenerate. However, future improvements in the
measurement of CP violation in $D-\overline D$ mixing will
start to probe alignment models.

\end{abstract}

\maketitle

\section{Introduction}

Measurements of flavor-changing neutral-current (FCNC)
processes in the quark sector put strong constraints on New
Physics (NP) at the TeV scale and provide a crucial guide for
model building. Generically, NP models can avoid existing
bounds by aligning the flavor structure with one of the quark
Yukawa matrices. However, new flavor breaking sources involving
only the $SU(2)_L$ doublet quarks $Q_i$ (i.e., breaking only the
$SU(3)_Q$ quark flavor symmetry) cannot be simultaneously
diagonalized in both the up and the down quark mass bases,
and new contributions to FCNCs are necessarily generated. To
constrain such models of flavor alignment, processes involving
both up and down type quark transitions need to be measured.
Consequently, one would na\"ively conclude that robust
constraints on the corresponding microscopic flavor structures
come from the {\it weaker} of the bounds in the
up and the down sectors.

Below we argue, however, that in a large class of models,
contrary to flavor violation in $\Delta F=2$
processes~\cite{Blum:2009sk}, in the case of $\Delta F=1$
CP violation, it is the {\it strongest} of the up and down sector
constraints which applies. We show that in these
scenarios, to a good approximation, the sources of $\Delta F=1$
CP violation are universal, namely they do not transform under
flavor rotations between the up and the down mass bases. This
is particularly important for the NP interpretation of the
recent LHCb evidence for CP violation in $D$ decays. Employing
the $\epsilon'/\epsilon$ constraint on new CP violating $\Delta
s=1$ operators, we exclude sizable contributions of $SU(3)_Q$
breaking NP operators to the direct CP asymmetries in
singly-Cabibbo-suppressed $D$ decays, in particular to $\Delta
a_{CP}$ measured by the LHCb experiment~\cite{Aaij:2011in}.

Furthermore, applying our argument to rare semileptonic $K$ and
$B$ decays, we show how the present and future measurements of
these processes constrain the sources of CP violation in rare
semileptonic $D$ decays and FCNC top decays. In particular, the
observation of non-SM CP asymmetries in these processes would,
barring cancellations, signal the presence of new sources of
$SU(3)_{U,D}$ flavor symmetry breaking.

Finally, an additional implication of our result is that in
viable flavor alignment models the universal flavor and CP
violating phases are naturally small. Applying this insight to
supersymmetric (SUSY) alignment models leads to the conclusion that
the first two generation squarks can have mass splittings as
large as 30\% at the TeV scale, consistent with mass anarchy at
a supersymmetry breaking mediation scale as low as 10~TeV.

\section{Universal contributions to CP Violation with Two Generations}
\label{sec:2g}

It is well known that the gauge sector of the Standard Model
(SM) respects a large global flavor symmetry. In the quark
sector, the corresponding flavor group, ${\cal G}_F =
SU(3)_Q\times SU(3)_U \times SU(3)_D\,$, is broken by the up
and the down Yukawa matrices $Y_{u,d}\,$, formally transforming
as $(3,\bar 3,1)$ and $(3,1,\bar 3)$ under ${\cal G}_F\,$,
respectively. From these, one can construct two independent
sources of $SU(3)_Q$ breaking,
\beq
\au \equiv \yut \,, \qquad \ad \equiv \ydt \,,
\eeq
which both transform as $(8,1,1)$ under ${\cal G}_F\,$, where
$\slashed{\mathrm{tr}}$ denotes the traceless part. Our
argument applies to all NP flavor structures, $X$, which can be
written in the form
\beq \label{xq_operator}
\co_L = \Big[(X_L)^{ij}\, \overline Q_i \gamma^\mu Q_j \Big] L_\mu \,.
\eeq
Here $Q_i$ stands for the left-handed quark doublets, $i$ and
$j$ are generation indices and $X_L$ is a traceless Hermitian
flavor matrix. The $L_\mu$ denotes a flavor-singlet current,
such as
\beq
L_\mu = \sum_q \overline q \gamma_\mu q \,, \qquad
  \sum_\ell \overline \ell \gamma_\mu \ell \,, \qquad
  H^\dagger D_\mu H \,, \quad \ldots\,,
\eeq
that is, a sum over quarks or leptons, a Higgs current, etc.
Note that the way that color and spinor indices are contracted
in Eq.~\eqref{xq_operator} is irrelevant for our argument
below.

It is easy to show that in the two generation limit, the unique
Jarlskog invariant relevant for $\Delta F=1$ CP violation
sourced by $\co_L$ ($X_L$) is proportional to $X_L^{CPV}\equiv
\tr( X_L \cdot J)$~\cite{arXiv:1002.0778}, where
\beq
J\equiv i[\au,\, \ad]\,.
\eeq
The expression for $X_L^{CPV}$ is manifestly
reparametrization invariant and thus basis independent. A
nontrivial feature of such $SU(2)_Q$ breaking is that
$\au\,,\, \ad\,,\, J$ form a complete basis in the
three-dimensional space of traceless Hermitian $2\times 2$
matrices, and that $J$ is orthogonal to the other two directions,
i.e., ${\rm Tr} (\mathcal A_{u,d} \cdot J) = 0$. It follows
that the imaginary (CP violating) part of $X_L\,$, proportional
to $X_L^{CPV}$, is also orthogonal to the plane of
$\au,\, \ad\,$. It is thus invariant under flavor rotations in
this plane and in particular under the two-dimensional real CKM
rotation between the up and the down quark mass bases.
Consequently, the amount of CP violation generated by $X_L$ in
both up and down quark transitions is the same, irrespective of
the direction of the projection of $X_L$ in the $(\au,\, \ad)$
plane.

Explicitly, the up and down quark components of $\mathcal O_L$
in their relevant mass bases are
\beq
\left[(X^u_L)_{ij}\, \bar u_L^i \gamma^\mu u_L^j\right] L_\mu
\,, \qquad \left[(X^d_L)_{ij}\, \bar d_L^i \gamma^\mu d_L^j
\right] L_\mu\,,
\eeq
where $X^{u,d}_L$ are $X_L$ rotated to the up and down mass
bases, respectively. The universality of CP violation in
$\Delta F=1$ transitions can now be expressed explicitly as
\beq\label{Uni2gen}
\im(X^u_L)_{12} = \im(X^d_L)_{12}\propto \tr \left( X_L \cdot
J\right)\, .
\eeq
Another simple way to understand this universality of CP
violation is to notice that in the up or down mass basis, $J$
is proportional to the Pauli matrix $\sigma_2\,$, which is
invariant under $SO(2)$ rotations. A consequence of
Eq.~\eqref{Uni2gen} is that CP violation in both the up and the
down sectors vanishes if $X_L$ is in the plane of $\au$ and
$\ad$ (and in particular if $X_L$ is aligned with $\au$ or
$\ad$), as can also be seen from the vanishing of the Jarlskog
invariant. The difference compared to $\Delta F=2$ flavor
violation follows from the fact that in the latter case CP
violation is proportional to $\im\big[(X^{u,d}_L)^2\big] = 2
\,\im X^{u,d}_L\, \re
X^{u,d}_L$~\cite{Blum:2009sk,arXiv:1002.0778}, which depends
also on the non-universal real part. In addition, many CP
violating observables also depend on hadronic matrix elements,
which modify the universal short distance contributions we
focus on, but do not introduce dependence on any new
invariants.

The two-generation limit can only be considered as approximate
for the strange and charm sectors. Furthermore, it is not
immediately clear how it can be relevant for $\Delta F=1$
transitions involving the third generation quarks. We address
these two issues in turn. In both cases we use the fact that
the SM possesses an approximate $SU(2)_Q$ flavor symmetry,
which is broken only by $(m_{c,s}^2-m_{u,d}^2)/m_{t,b}^2$ and
the $\theta_{13}$ and $\theta_{23}$ CKM mixing angles.

\section{Universal contributions to CP Violation with Three Generations}
\subsection{CP violation involving the first two generations within the three flavor framework}

To describe $\Delta c,\, \Delta s=1$ processes in the context
of three generations, we can decompose $X_L$ according to the
$SU(2)_Q$ symmetry. Taking advantage of the SM $SU(2)^3$
symmetry obtained when the first two generation masses are
neglected~\cite{GMFV}, one can choose a basis which isolates
the large eigenvalues in the up and down Yukawa matrices. These
become block diagonal in the limit where the small CKM mixing
angles $\theta_{13}$ and $\theta_{23}$ are neglected. In this
basis, $(X_L)_{33}$ does not transform under the $SU(2)_Q$
symmetry. The upper 2$\times$2 block of $X_L\,$, which we
denote $(X_L)_2\,$, transforms as an adjoint of $SU(2)_Q$
corresponding to the two-generation case discussed above. In
addition, $X_L$ consists of an extra doublet of $SU(2)_Q\,$,
which we denote $x_L\,$, composed of $(X_L)_{13}$ and
$(X_L)_{23}\,$. At leading order, $x_L$ only mediates $\Delta
b,\, \Delta t=1$ processes, while $\Delta c,\, \Delta s =1$
processes can be generated at order $x_L^2\,$. Thus, we expect
its contributions to $\Delta c,\, \Delta s = 1$ processes to be
subdominant and also generically independent of those of
$(X_L)_2\,$. In practice, our result in Eq.~\eqref{Uni2gen}
applies separately to contributions from $(X_L)_2$ and
$x_L^2\,$, barring cancellations. Further corrections to
Eq.~\eqref{Uni2gen} come from the SM breaking of the
$SU(2)_Q\,$, but these are suppressed by powers of
$m_c^2/m_t^2$ or $m_s^2/m_b^2\,$.

\subsection{CP violation in third generation transitions}

The universality of CP violation also holds for flavor
transitions involving third generation quarks. A useful
approximation is again to neglect the masses of the first two
generation quarks. The breaking of the flavor symmetry by
$Y_{u,d}$ is then characterized by $SU(3)/SU(2)$~\cite{GMFV}.
In this limit, the 1--2 rotation and the phase of the CKM
matrix become unphysical, and we can, for instance, apply an
$SU(2)$ rotation to the first two generations to ``undo" the
1--3 rotation. Therefore, the CKM matrix is effectively reduced
to a real matrix with a single rotation angle $\theta \simeq
\sqrt{\theta_{13}^2 + \theta_{23}^2}$ between an ``active"
light flavor, $q_a\,$, and the 3rd
generation~\cite{arXiv:1002.0778}. Such a pattern of flavor
breaking respects an approximate $U(1)_Q$ symmetry for the
combination of light quarks that effectively decouples, thus
ensuring that all interactions of this ``sterile" flavor,
$q_s\,$, are CP conserving. What remains is an effective
two-generation system with a single measure of CP violation in
transitions between the 3rd generation and the active light
flavor. It is given again by ${\rm Tr} (X_L\cdot J)$, which is
flavor basis independent, and thus universal
(see~\cite{Gedalia:2010rj} for an extended discussion).
Therefore, to leading order, there is a universal relation for
CP violation involving transitions between the third generation
and the up and down component of the active states,
\beq
\im (X_L^u)_{a3} = \im (X_L^d)_{a3}\,, \label{3rdgen1}
\eeq
where the active states coincide with the second generation
quarks up to ${\cal O}\left(\lambda_C
\right)$~\cite{arXiv:1002.0778}, where $\lambda_C\simeq 0.23$
is the sine of the Cabibbo angle.

The leading corrections to Eq.~\eqref{3rdgen1}, in the massless
two-generation limit, can be understood by decomposing $X_L$ to
its representations under the $SU(2)$ 3rd generation--active
flavor group, $SU(2)_{3a}\,$. Besides the adjoint contribution
of $(X_L)_{a3}\,$, the entries $(X_L)_{sa}$ and $(X_L)_{s3}$
form an $SU(2)_{3a}$ doublet. At order $(X_L)^2$, they in term
produce a new adjoint of $SU(2)_{3a}\,$, which would induce an
independent contribution to the transitions between the 3rd
generation and the active flavor, and hence correct the
relation in Eq.~\eqref{3rdgen1}. Since these are independent
contributions, barring cancellations, this relation would still
hold for each of them separately. Furthermore, the extra
$SU(2)_{3a}$ doublet would in general contribute to transitions
between the first two generations, and  should therefore be
strongly constrained. We thus conclude that we expect the
expression in Eq.~\eqref{3rdgen1} to hold to a good accuracy.

Finally, we comment on the fact that one can constrain third
generation alignment with data involving the first two
generations. Consider, for instance, an alignment model that
saturates the bounds from $B_d$ mixing, including CP violation.
In other words, the TeV-scale new physics contributions are
required to be approximately aligned with the down Yukawa. This
structure would necessarily contribute to CP violation in
$D-\overline D$ mixing, since the real and the imaginary parts
cannot be simultaneously eliminated. Such a scenario was
investigated in~\cite{arXiv:1002.0778}, where it was shown that
in practice the resulting contributions are still two-to-three
orders of magnitude below the present bounds.

\section{Examples}

\subsection{\boldmath Relating CP violation in hadronic $K$ and $D$ decays}

The argument presented in Sec.~\ref{sec:2g} allows us to relate
the existing constraints in the kaon system to $SU(3)_Q$
breaking NP contributions to direct CP violation in the charm
system, and in particular, to relate $\Delta a_{CP}$ to
$\epsilon'/\epsilon$. The relevant $SU(3)_Q$ breaking NP
operators in Eq.~\eqref{xq_operator} induce at low energies
contributions of the form $Q^q_{1,2,5,6}$ defined in Eqs.~(8)
and~(15) of~\cite{Isidori:2011qw}. The weakest bound on any of
these operators from $\epsilon'/\epsilon$ is given
by~\cite{Isidori:2011qw}
\beq \label{epsilon_prime_bound}
\big|\im (C_2^{(0)}) \big| \lesssim 4.5 \times 10^{-5}\,\left(
\frac{\Lambda_{\rm NP}}{350\,\mathrm{GeV}} \right)^2 ,
\eeq
for $Q_2^{(0)}=(\bar d_\alpha s_\beta)_{V-A}\, \sum_q (\bar
q_\beta q_\alpha)_{V-A}\,$, where $\alpha$ and $\beta$ are
color indices, and the sum over $q$ includes the $u,d,s,c,b$
flavors. The contributions of such operators to $\Delta a_{CP}$
are given by~\cite{Isidori:2011qw}
\beq \label{delta_acp}
\Delta a^{\rm NP}_{CP} \approx 8.9 \sum_i \im(C^{\rm NP}_i)\,
\im(\Delta R^{\rm NP}_i) \,,
\eeq
where $\Delta R^{\rm NP}_i$ denotes the ratio of the NP
amplitude and the leading SM ``tree" contribution. Applying the
bound in Eq.~\eqref{epsilon_prime_bound} to
Eq.~\eqref{delta_acp}, and assuming $\Delta R^{\rm NP}_i\sim
1$, we find
\beq
\Delta a^{\rm NP}_{CP} \lesssim 4 \times 10^{-4} \,.
\eeq
We thus learn that in any $SU(2)_L$ invariant NP model,
the contributions of the $Q^q_{1,2,5,6}$ operators to
$\Delta a_{CP}$ must be negligible.

\subsection{\boldmath Semileptonic $K$ and $D$ decays}

An important class of rare $K$ decays are those involving a
pion and a lepton pair. The short distance contributions to
$K_L \to \pi^0 \ell^+ \ell^-$ or for $K_L \to \pi^0
\nu \bar \nu$ are dominantly CP violating.  (The
$K_L \to \pi^0 \ell^+ \ell^-$ decay also receives a
non-negligible CP conserving long distance contribution.)
So far, only upper bounds on these rates have been set~\cite{PDG}
\beq \label{semilep_exp}
\begin{split}
\br &(K_L \to \pi^0 e^+ e^-)<2.8 \times 10^{-10} \,, \\ \br &
(K_L \to \pi^0 \mu^+ \mu^-)<3.8 \times 10^{-10} \,,
\\ \br &(K_L \to \pi^0 \nu \bar \nu)<2.6 \times 10^{-8}
\,,
\end{split}
\eeq
all at 90\% confidence level. These experimental results can
be translated into constraints on the Wilson coefficients
of the appropriate effective operators:
\beq \label{semilep_hamil}
\mathcal{H}^{\rm eff}_{\Delta s=1} \supset
\frac{C^{\ell_{R/L}}_{sd}}{\Lambda_{\rm NP}^2} (\bar s d)_{V-A}\,
(\bar \ell \ell)_{V \pm A}+ \frac{C^\nu_{sd}}{\Lambda_{\rm NP}^2}
(\bar s d)_{V-A}\, (\bar \nu \nu)_{V-A} \,,
\eeq
where $\Lambda_{\rm NP}$ is the NP scale and the superscripts $\ell_{R/L}$
distinguish the operators containing the right handed ($V+A$) and left
handed ($V-A$) charged lepton currents (in the standard notation
$(C^{\ell_R}_{sd} \pm C^{\ell_L}_{sd})/2$ are $C_{9,10}$).

We start by analyzing the process $K_L \to \pi^0 \ell^+
\ell^-$. Following~\cite{hep-ph/0606081, arXiv:0912.1625}, we
can neglect the SM contributions, which are of order 10\% or
less compared to the current experimental limits. Taking the
central values for all the parameters entering the theoretical
prediction from~\cite{hep-ph/0606081} and comparing with
Eq.~\eqref{semilep_exp}, we obtain the constraints
\beq \label{cl_bound}
\begin{split}
|\im \, C^{e_{R/L}}_{sd}| &< 5.5 \times 10^{-4}\,\left(
\frac{\Lambda_{\rm NP}}{1\,\mathrm{TeV}} \right)^2 , \\
|\im \, C^{\mu_{R/L}}_{sd}| &< 9.5 \times 10^{-4}\,\left( \frac{\Lambda_{\rm
NP}}{1\,\mathrm{TeV}} \right)^2 .
\end{split}
\eeq

Similarly, we consider the decay channel involving neutrinos.
Since the class of operators in Eq.~\eqref{xq_operator}
conserves lepton flavor, we can use the Grossman-Nir bound
(instead of the presently weaker experimental bound), which
relates the rates for charged and neutral kaon
decays~\cite{hep-ph/9701313}:
\beq \label{gn}
\br (K_L \to\pi^0 \nu \bar \nu)<4.4\, \br (K^+ \to\pi^+ \nu \bar
\nu)\,.
\eeq
The latter branching ratio is~\cite{PDG}
\beq
\br (K^+ \to\pi^+ \nu \bar \nu)=(1.73^{+1.15}_{-1.05})\times
10^{-10} \,.
\eeq
Taking a 90\% confidence level upper bound and comparing it
with the theoretical predictions,
following~\cite{arXiv:0912.1625}, we obtain
\beq \label{cnu_gn_bound}
|\im \, C^\nu_{sd}|< 2.6 \times 10^{-4}\,\left(
\frac{\Lambda_{\rm NP}}{1\,\mathrm{TeV}} \right)^2 \,.
\eeq

Due to our CP violation universality argument, the bounds in
Eqs.~\eqref{cl_bound} and~\eqref{cnu_gn_bound} apply directly
to the charm system as well. The appropriate observables are CP
asymmetries involving rare $D$ semileptonic decays, for
example:
\beq \label{Dasymmetry}
a^{D}_e \equiv \frac{\br(D^+ \to \pi^+ e^+ e^-)- \br(D^- \to
\pi^- e^+ e^-)}{\br(D^+ \to \pi^+ e^+ e^-)+ \br(D^- \to \pi^-
e^+ e^-)}\,,
\eeq
as well as for neutrinos instead of charged leptons in the final
state (see, e.g.,~\cite{Paul:2011ar}). An
upper bound on the asymmetry in Eq.~\eqref{Dasymmetry} can be
obtained as follows. We assume that the SM contribution is
essentially CP conserving, so that CP violation is dominated by NP, and
that the overall decay rate is dominated by long distance
SM contributions~\cite{hep-ph/0112235}. Denoting the NP and the SM
amplitudes as $\mathcal A_{\rm NP} $ and  $\mathcal A_{\rm
SM} $ respectively, with $|\mathcal A_{\rm NP}| \ll |\mathcal
A_{\rm SM}|$, we can write
\beq
\left|a^{D}_e\right| \lesssim \frac{2\int d\rho\,
  |{\rm Im}(\mathcal A_{\rm NP})|\, |\mathcal A_{\rm SM}| }
  {\int d\rho\, |\mathcal A_{\rm SM}|^2}
\lesssim 2\, \sqrt{\frac{\int d\rho\, |\mathcal A_{\rm NP}|^2}
  {\int d\rho\, |\mathcal A_{\rm SM}|^2}} \,,
\eeq
where $\int d\rho$ denotes the relevant three body phase space
integration. We can identify the denominator of the right-hand
side with the square root of the {\it experimentally
determined} rate $\Gamma(D^+ \to \pi^+ e^+ e^-)$, avoiding the
theoretically uncertain evaluation of the SM long distance
amplitude. On the other hand, the numerator is dominated by
short distance physics, and can be computed reliably using the
recent lattice QCD calculation of the $D\to \pi$ form
factor~\cite{Na:2011mc}, yielding
\beq \label{Dasym_bound}
\left|a^{D}_e\right| \lesssim \left(\frac{1\,\rm
TeV}{\Lambda_{\rm NP}}\right)^2 \frac{0.1\, |\im \,
C^{e_{R/L}}_{sd}|}{\sqrt{\br(D^+ \to \pi^+ e^+ e^-)}} \lesssim
0.02\,.
\eeq
On the right-hand side we used Eq.~\eqref{cl_bound} and the
experimental upper bound on the branching ratio~\cite{PDG},
given that it is close to the estimated long distance
contributions~\cite{hep-ph/0112235}. If the above bound would
be experimentally violated, the source of the required CP
violation could not be of the form of Eq.~\eqref{xq_operator}.
Finally, we note that this constraint may be refined in the
future with improved experimental bounds and theoretical
estimates of the relevant processes in either the $K$ or the
$D$ systems.

\subsection{\boldmath Semileptonic $B$ decays}

Rare semileptonic $B$ decays $B\to X_s \ell^+ \ell^-$, $B\to
K^{(*)} \ell^+ \ell^-$, $B_s \to \mu^+\mu^-$ and $B\to K^{(*)}
\nu\bar\nu$ offer direct probes of NP contributions of the form
of Eq.~\eqref{xq_operator}. At the moment, the most sensitive
probe of this kind of NP contribution is the partial branching
ratio of the inclusive decay $B\to X_s \ell^+ \ell^-$ in the
so-called ``low-$q^2$ region", $q^2 \equiv (p_{\ell^+} +
p_{\ell^-})^2 \in [1,6]\, \mathrm{GeV}^2$.~\footnote{We checked
that other related presently measured and theoretically clean
observables like the low-$q^2$ forward-backward asymmetry ($A_{FB}$) in $B\to K^*
\ell^+\ell^-$~\cite{AFBLHCb}, the high-$q^2$ region in $B\to
X_s\ell^+\ell^-$~\cite{Ligeti:2007sn, Huber:2007vv}, or the
leptonic decay $B_s \to \mu^+\mu^-$~\cite{BsmumuLHCbCMS} do not
yield competitive bounds on the NP contributions that we
consider here.} The operator in Eq.~\eqref{xq_operator}
contributes to the effective weak Hamiltonian, similar to
Eq.~(\ref{semilep_hamil}),
\begin{align}
\mathcal H^{\rm eff}_{\Delta b =1} \supset
\frac{C^{\ell_{R/L}}_{bs}}{\Lambda_{\rm NP}^2}\, (\bar b s)_{V-A}\, (\bar \ell \ell)_{V\pm A}
+\frac{C^{\nu}_{bs}}{\Lambda_{\rm NP}^2}\, (\bar b s)_{V-A}\, (\bar\nu \nu)_{V-A}
\,.
\label{eq:bsops}
\end{align}
Employing the relevant semi-analytic NP formulae for both the
electron and muon channels~\cite{Huber:2005ig}, we can derive
bounds on $C^i_{\rm NP}\,$. The experimental results are
presented averaged over the electron and muon
channels~\cite{hep-ex/0404006}, resulting
in~\cite{Huber:2007vv}
\begin{equation}
\mathrm{Br}(B\to X_s \ell^+ \ell^-)_{\rm low}
  = (1.60\pm 0.50)\times 10^{-6}\,.
\end{equation}
To bound the operators in Eq.~\eqref{eq:bsops}, we require that
the NP contribution to the particular leptonic channel should
be consistent with the above averaged value. In order to
extract robust bounds on $\im(C^i_{\rm NP})$ from
$\mathrm{Br}(B\to X_s \ell^+ \ell^-)_{\rm low} $, we
marginalize over the corresponding real parts as well as the SM
theoretical uncertainties as given in~\cite{Huber:2005ig}. In
this way we obtain at $95\%$ C.L.\footnote{Recent
analyses~\cite{Altmannshofer:2011gn} of NP in semileptonic
$b\to s$ transitions obtained somewhat stronger bounds by
relying on high-$q^2$ $A_{FB}$ and partial branching ratio
measurements of exclusive $B\to K^{(*)} \ell^+ \ell^-$ decays.
We do not consider these observables, since they are subject to
substantial theoretical (form factor) uncertainties.}
\begin{align}\label{bsllbounds}
\big|\mathrm{Im}(C^{\ell_L}_{bs})\big| & <  1.6 \times 10^{-3}
  \left(\frac{\Lambda_{\rm NP}}{1\,\rm TeV}\right)^2 , \qquad \mathrm{for~} \ell = e,\mu\,, \nonumber\\
\big|\mathrm{Im}(C^{\ell_R}_{bs}) \big| & <  8.5 \times 10^{-4}
  \left(\frac{\Lambda_{\rm NP}}{1\,\rm TeV}\right)^2 , \qquad \mathrm{for~} \ell = e,\mu \,.
\end{align}
Finally, $C^{\nu}_{\rm NP}$ can be bounded directly from the
experimental searches for the $B\to K^{(*)}\bar \nu\nu$
decays~\cite{BKnunuExp}, which yield~\cite{Kamenik:2011vy}
\begin{equation}
\big|\mathrm{Im}(C^{\nu}_{bs})\big| < 7.5 \times 10^{-3}
  \left(\frac{\Lambda_{\rm NP}}{1\,\rm TeV}\right)^2, \qquad {\rm for~all}~\nu\,.
\end{equation}

A similar analysis could in principle be performed also for the
$b\to d$ transitions. However, at present, the associated
experimental constraints are much weaker~\cite{PDG}, and no
interesting bounds can be obtained.

In the long run, the strongest constraints on NP contributions
with a new weak phase to the $C^{\ell_{R/L}}_{bs}$ Wilson
coefficients in Eq.~(\ref{eq:bsops}) (again $(C^{\ell_R}_{bs}
\pm C^{\ell_L}_{bs})/2$ are $C_{9,10}$ in the rare $b$ decay
literature) may come from CP violation studies in $b\to
s\ell^+\ell^-$ mediated decays. These operators dominate in the
large-$q^2$ region, while the electromagnetic penguin operator,
$O_7\,$, is also important at small $q^2$. The $B\to K^{*}
\ell^+ \ell^-$ mode is particularly promising, since the
distribution of the $K^* \to K\pi$ decay products allows to
extract information about the polarization of the $K^*$. When
combined with the angular distributions of the two charged
leptons, it is possible to construct observables probing
directly CP violating contributions to the relevant
short-distance Wilson coefficients~\cite{CPVinBKll}. Such
observables could potentially be measured at LHCb and
Super$B$~\cite{SuperB}. On the other hand, the direct CP
asymmetries depend on strong phases, which are small in the
inclusive $B\to X_s\ell^+\ell^-$ decay (outside the resonance
region), and are poorly known in the exclusive $B\to
K^{(*)}\ell^+\ell^-$ case. Another probe of this physics could
be the study of time-dependent CP asymmetries in these modes.
While these are challenging experimentally, the interpretation
of the results would be theoretically cleaner. The SM predicts
that the time-dependent CP asymmetry vanishes, as it does in
$B_s\to \phi\phi$, to an even better accuracy than in $B_s\to
\psi\phi$, due to a $2\beta_s - 2\beta_s$ cancellation between
the mixing and decay phases. The same cancellation occurs in NP
models in which the mixing amplitude is modified as
$M_{12}^{\rm SM} \times R^2$ and the decay amplitude is
modified as $A^{\rm SM} \times R$. While this is the case in
most supersymmetric models, it is not generic, and is violated,
for example, by models containing a $Z'$ which has a flavor
changing coupling to quarks and non-universal couplings to
quarks and leptons. (With very large data sets at the upgraded
LHCb, a time-dependent $B_s\to \mu^+\mu^-$ analysis would also
be worth pursuing.)

To analyze the connection between $t\to cZ$ and FCNC $b\to s$
decays, we need to consider the NP operators before the $Z$ is
integrated out~\cite{Fox:2007in}. For example, the operator
$(\bar b s)_{V-A}\, (H^\dagger DH)$ contributes to
Eq.~(\ref{eq:bsops}), since after electroweak symmetry breaking
$H^\dagger D_\mu H \to g v^2 Z_\mu$. Thus the relevant Wilson
coefficient, $C^{H}_{bs}\,$, is constrained from $B\to
X_s\ell^+\ell^-$, similar to Eq.~(\ref{bsllbounds}), as
$\big|\mathrm{Im}(C^H_{bs}) \big| <  8.7 \times 10^{-3}\,
(\Lambda_{\rm NP}/{\rm TeV})^2$. Top decays into final states
with a jet and a pair of charged leptons offer a probe of the
related $(X^u_L)_{tc}$ and $(X^u_L)_{tu}$
contributions~\cite{Drobnak:2008br}. The expected sensitivity
of this mode with 100\,fb${}^{-1}$ at the 14~TeV LHC is
$|C^H_{tc(u)}| \lesssim 0.2\, (\Lambda_{\rm NP} / {\rm
TeV})^2$~\cite{Carvalho:2007yi,Fox:2007in}, where the relevant
operator is defined as $(\bar t c(u))_{V-A}\, (H^\dagger DH)$.
According to Eq.~\eqref{3rdgen1}, we can conclude that barring
cancellations, any experimental signal of CP violation in this
channel would have to be due to $SU(3)_U$ breaking NP.

\section{Implications for SUSY models}

In SUSY models the left-handed squark mass-squared
matrix, $\tilde m_Q^2\,$, is the only source of $SU(3)_Q$
breaking, and is approximately $SU(2)_L$ invariant~(see,
e.g.,~\cite{Nir:2007xn} and references therein). In the
following we discuss a universal constraint on $\tilde m_Q^2$
from $\Delta F=1$ CP violation. In addition, we consider an
example of $\Delta F=2$ constraints in relation to alignment
models, where our argument about universality of the CP phase
also plays a role. In all cases the bounds can be directly
applied on the corresponding mass insertion parameters.

First we analyze the constraint from $\epsilon'/\epsilon$. In
the super-CKM basis, the neutral gaugino couplings are flavor
diagonal, while the mass matrices of the squarks are not
diagonal in general. New contributions to CP violation in
$\Delta F=1$ processes involving left handed quarks are induced
by the imaginary off-diagonal elements of $\tilde m_Q^2\,$, and
can be parameterized in terms of the ratios $\delta^{ij}_{LL}
\equiv {\left(\tilde m_Q^2\right)^{ij} \! /\, \bar m_{\tilde
Q}^2 } \,,$ where $i,j=1,2$ are flavor indices and $\bar
m_{\tilde Q} \equiv (m_{\tilde Q_1}+m_{\tilde Q_2})/2$ is the
average squark mass (this choice is consistent to linear order
with the convention of~\cite{Gabbiani:1996hi}). The
experimental constraint on new contributions to
$\epsilon'/\epsilon$ is translated to the following bound on
the left-handed mass insertion parameter~\cite{Gabbiani:1996hi}
$\im\, \delta^{12}_{LL}\leq 0.5$ for $\bar m_{\tilde Q}
  = m_{\tilde g}= 500~\rm GeV\,.
$ This can be straightforwardly rephrased as a robust constraint
on the level of degeneracy
\beq
\delta^{12}_{Q} \equiv \frac{m_{\tilde Q_2}-m_{\tilde
Q_1}}{m_{\tilde Q_2}+m_{\tilde Q_1}} \le 0.25 \left(\frac{\rm
500\,GeV}{\bar m_{\tilde Q}}\right) .
\eeq
This bound is weaker than the one obtained by combining the
bounds from $\epsilon_K$ and $D-\overline D$
mixing~\cite{Blum:2009sk}. Yet, interestingly, it could have
constrained degeneracy without the need for any additional
measurements involving $D$ mesons, more than 20 years ago
already, when the experimental uncertainty of
$\epsilon'/\epsilon$ approached the $10^{-3}$
level~\cite{Burkhardt:1988yh}.

\begin{figure}[tb]
  \centering
  \includegraphics[width=.6\textwidth]{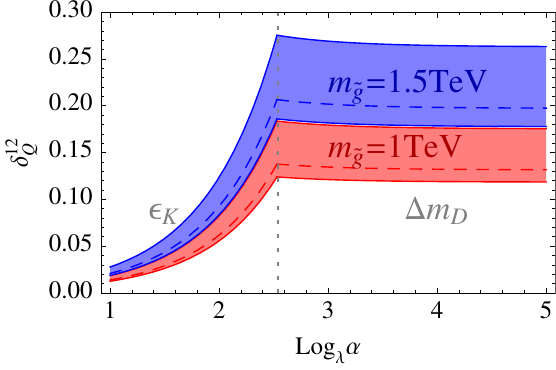}
  \caption{The bound on $\delta_Q^{12}$ as a function of the angle $\alpha$ (see
  text). The angle $\alpha$ is plotted on a log scale in the basis $\lambda_C=
  0.23$, so that a value of 1 on the $x$ axis corresponds to $\alpha=\lambda_C$
  (large angle), while a value of 5 gives $\alpha=\lambda_C^5$ (small angle ---
  down alignment). The vertical doted line shows the angle of optimal alignment
  (weakest bound). The red (blue) shaded region corresponds to a gluino mass
  $m_{\tilde g}$ of 1~(1.5)~TeV, and inside each region the average squark mass
  $\bar m_{\tilde Q}$ is varied in the range [$0.8\,m_{\tilde g},\, 1.2\,m_{\tilde g}$].
  The upper edge of each region (weakest bound) comes from the lowest $\bar
  m_{\tilde Q}\,$. The two dashed lines correspond to $\bar
  m_{\tilde Q}=m_{\tilde g}\,$.
  \label{fig:DDvsKK}}
\end{figure}

Constraints on alignment models that balance the bounds from
mixing and CP violation in the $K$ and $D$ systems have been
analyzed in~\cite{Blum:2009sk}. Here we comment on their
results for supersymmetric models based on our CP universality
argument. According to the parameterization employed
in~\cite{Blum:2009sk}, $\sin \alpha$ ($\sin 2\gamma$) is
proportional to the real (imaginary) part of the off-diagonal
element of the NP flavor violating source in the down mass
basis. CP universality implies that in the up mass basis, $\sin
2\gamma$ still corresponds to the imaginary part, while the
real part is rotated by twice the Cabibbo angle. Equation~(31)
in~\cite{Blum:2009sk} gives the bounds on squark mass
degeneracy for the cases of vanishing ($\sin 2\gamma=0$) and
maximal ($\sin 2\gamma \sim 1$) phase. We argue that the latter
case is irrelevant, since it violates the assumption of
alignment. In contrast, while realistic models of alignment
generically do not control the fundamental CP violating phases,
they force both $\sin \alpha$ and $\sin 2\gamma$ to be small,
and should therefore be taken to be
comparable~\cite{Nir:1993mx}. This leads to a much weaker bound
than the more stringent one in~\cite{Blum:2009sk}. In
particular, the bound on $\delta_Q^{12}$ from $\epsilon_K$ and
$\Delta m_K$ for $\sin \alpha \sim \sin 2\gamma$ is shown in
Fig.~\ref{fig:DDvsKK} as a function of the angle $\alpha$, for
various ranges of the relevant SUSY parameters (see the
caption). It can be seen that on the right-hand side of the
plot, where the angle is very small (down alignment), the
strongest constraint comes from $\Delta m_D\,$, while on the
left hand side, where the angle is large, $\epsilon_K$ is the
dominant constraint. The vertical dashed line marks the
transition point, where the alignment is optimal, yet as
evident from the plot, making the angle smaller only mildly
affects the bound on $\delta_Q^{12}$. For the case where the
gluino mass and the average squark mass are both 1~TeV, the
weakest bound is $\delta_Q^{12} \lesssim 0.13$. This occurs
around $\log_\lambda\alpha \sim 2.5$, so the universal CP
violating phase is of order $\lambda_C^{2.5}$. This implies an
upper bound on CP violation in $D-\overline D$ mixing of order
$0.2$, around the current experimental limit on
$\big||q/p|-1\big|$~\cite{Asner:2010qj}, which is expected to
be improved significantly in the near future.

It is interesting that a modest level of degeneracy can be
obtained only from the renormalization group equation (RGE)
flow, when starting from anarchy at the SUSY breaking mediation
scale~\cite{Nir:2002ah}. Moreover, in order to satisfy the
bounds on degeneracy from optimal alignment models, as
presented in Fig.~\ref{fig:DDvsKK}, the mediation scale does
not have to be very high. To show this, we use the SUSY RGE for
the diagonal squark mass entries, which is dominated by the
gluino contribution. Neglecting the other gaugino
contributions, we can solve the relevant equations at one loop
analytically
\begin{align}
\frac{1}{\alpha_s(M_S)} & = \frac{1}{\alpha_s(\Lambda)}
  + \frac{b_3}{2\pi}\, \ln \frac{\Lambda}{M_S}\,,\\
\frac{ m_{\tilde g}(\Lambda)}{ m_{\tilde g}(M_S)} & = 1
  + \alpha_s(\Lambda)\, \frac{b_3}{2\pi}\, \ln \frac{\Lambda}{M_S}\,,\\
m_{\tilde Q_{1,2}}^2(M_S) - m_{\tilde Q_{1,2}}^2 & (\Lambda) =
  \frac{8}{3 b_3}\, \big[m_{\tilde g}(\Lambda)^2 -m_{\tilde g}(M_S)^2\big]\,,
\end{align}
where $\Lambda$ is the typical scale of the new supersymmetric
particles (taken to be 1~TeV), $M_S$ is the SUSY breaking
mediation scale, $b_3=-3$ is the MSSM QCD beta function and the
last equation is written in the squark mass basis. In addition,
we define $\sum m_{\tilde Q}^2(\mu) = m_{\tilde Q_1}^2(\mu)+
m_{\tilde Q_2}^2(\mu)$ and $\Delta m_{\tilde Q}^2(\mu) =
m_{\tilde Q_2}^2(\mu)- m_{\tilde Q_1}^2(\mu)$. Then in our
approximation, only $\sum m^2$ has a nontrivial RGE evolution,
while $\Delta m^2$ is invariant. Writing
\beq
\delta_{Q}^{12}(\mu) = \frac{\, \Delta m_{\tilde Q}^2(\mu) }{
\sum m_{\tilde Q}^2(\mu)\, \Big[ 1+ \sqrt{1-\left(\Delta
m_{\tilde Q}^2(\mu) / \sum m_{\tilde Q}^2(\mu)\right)^2}\,
\Big] } \,,
\eeq
we observe that only the denominator has a nontrivial RGE
evolution. Furthermore, in the IR, $\delta_{Q}^{12}$ approaches
the limit $\delta_Q^{12}  \approx \Delta m^2_{\tilde Q} / 2
\sum m^2_{\tilde Q}\,$.

\begin{figure}[tb]
  \centering
  \includegraphics[width=0.75\textwidth]{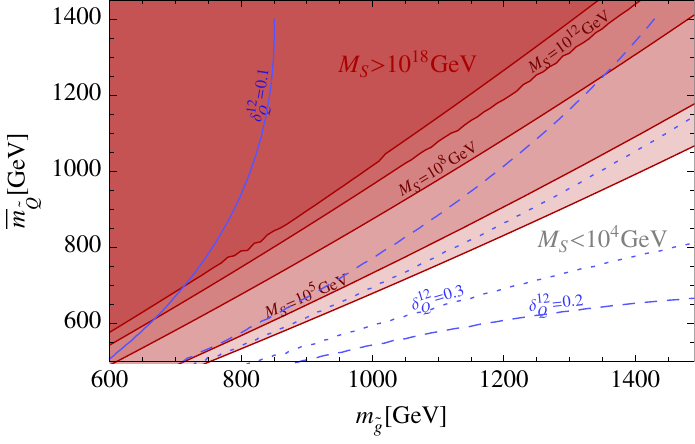}
  \caption{Contours for various values of the SUSY breaking mediation scale $M_S$ in the
  parameter space defined by $m_{\tilde g}$ and $\bar m_Q\,$, assuming
  $\delta_{Q}^{12}(M_S)=1$. Darker shaded regions correspond to higher
  $M_S\,$. Also shown are contours for $\delta_{Q}^{12}(\Lambda)=0.1,0.2,0.3$ in
  solid, dashed and dotted blue lines, respectively, where $\delta_{Q}^{12}(\Lambda)>0.3$
  between the two dotted lines.
  \label{fig:MS}}
\end{figure}

In Fig.~\ref{fig:MS} we show the contours of $M_S$ values that
yield the optimal $\delta_{Q}^{12}(\Lambda)$ as a function of
the gluino and average squark masses. Since $M_S$ is sensitive
to the level of anarchy assumed for $\delta_{Q}^{12}(M_S)$, we
choose conservatively to take $\delta_{Q}^{12}(M_S)=2$, which
actually corresponds to the extreme hierarchy case $m^2_{\tilde
Q_2} \gg m^2_{\tilde Q_1}$. Any finite ratio between the masses
would lead to a lower mediation scale than in
Fig.~\ref{fig:MS}. We find that, quite remarkably, a large
portion of the parameter space, with TeV superpartner masses,
is consistent with a fully anarchic spectrum at a moderate
mediation scale. Furthermore, we even find a non-negligible
region where  $\delta_{Q}^{12}(\Lambda)\sim 0.3$ is allowed.
For instance for a gluino mass of 1.3\,TeV we find that the
first two generation squark masses can be 550~GeV and 950~GeV
respectively at the TeV scale, which can hardly be considered
as a degenerate spectrum.

\section{Conclusions}

We have shown that NP that breaks the left handed $SU(3)_Q$
quark flavor symmetry induces approximately {\it universal}
contributions to CP violation in $\Delta F=1$ processes, in
that they are not affected by flavor rotations between the up
and the down mass bases. Therefore, these sources cannot be
aligned, and can be constrained by the strongest bound coming
either from the up or the down sectors. We have used this
result to show that the bound from $\epsilon'/\epsilon$
prohibits an $SU(3)_Q$ breaking explanation of the recent LHCb
evidence for CP violation in charm decays. A consequence of
this CP universality is that SUSY alignment models, even with a
low SUSY breaking mediation scale, are consistent with current
data, since the universal CP phase tends to be suppressed.
Therefore, fundamentally squarks need not be degenerate. We
note in this respect that the current direct experimental
searches for squarks are assuming degeneracy in the first two
generations, and therefore their lower bounds do not strictly
apply in the context of alignment models which could have a
significant splitting between the first two generations.
Finally, in this framework CP violation in $D-\overline D$
mixing is bounded from above with a maximal value which is
close to the current experiment sensitivity. Other types of
models of alignment (see~\cite{others}, for example), as in the
case of the SUSY example, also tend to yield more anarchy in
the up sector. Hence, they are expected to be constrained by
flavor transition measurements in the up sector, with the
contributions to CP violation somewhat suppressed.

We have also discussed the universality of CP violation
involving the third generation, and established a linkage
between CP violation in rare bottom and top quark decays, which
might be tested in the far future. It is interesting to note
that the combination of the current direct constraints on the
superpartner spectrum and naturalness implies the possibility
that the first two generation squarks are rather heavy, while
the third generation left handed squarks are approximately
degenerate~\cite{lightstops}. In such a case, flavor violation
involving the third generation would approximately satisfy our
universality condition. In this setup there is no generic
reason to expect the entries of the left-handed squark matrices
to be real. Thus, since the spectrum is hierarchal, the
experimental bound on the level of flavor violation can be
applied directly as constraints on the phases, which should be
of order $0.05$ (0.15) for $\Delta b=2$ processes in the $B_d$
($B_s$) systems for the third generation doublet and gluino of
around 500~GeV~\cite{Isidori:2010kg}. This further implies,
within this framework, a strong suppression of CP violating
processes involving only left handed squarks, in either the
down or the up sectors.

\begin{acknowledgments}
We thank Yuval Grossman, Gino Isidori, Yossi Nir and Michele
Papucci for useful discussions. We thank David Straub and
Wolfgang Altmannshofer for pointing out a numerical error in
Eq.~(22) in the first version of the manuscript. The work of
JFK was supported in part  by the Slovenian Research Agency. GP
is supported by the GIF, Gruber foundation, IRG, ISF and
Minerva. ZL was supported in part by the Office of High Energy
Physics of the U.S.\ Department of Energy under contract
DE-AC02-05CH11231.

\end{acknowledgments}

\end{document}